\documentclass{PoS}

\usepackage{graphicx}
\usepackage{amsmath}
\usepackage{xspace}

\newcommand{\mae}[3]{\langle#1\rvert#2\rvert#3\rangle}
\newcommand{\Mae}[3]{\bigl\langle#1\bigr\rvert#2\bigr\rvert#3\bigr\rangle}
\newcommand{\abs}[1]{\lvert#1\rvert}
\newcommand{\ord}[1]{\mathcal{O}(#1)}
\newcommand{\vev}[1]{\langle #1 \rangle}
\newcommand{\Vev}[1]{\bigl\langle #1 \bigr\rangle}
\newcommand{\braket}[2]{\langle#1\rvert#2\rangle}

\newcommand{\eq}[1]{Eq.~\eqref{eq:#1}}

\newcommand{\df}{\mathrm{d}}
\newcommand{\img}{\mathrm{i}}
\newcommand{\tr}{\textrm{tr}}
\newcommand{\sdt}{\!\cdot\!}

\newcommand{\ve}{\varepsilon}
\newcommand{\eps}{\epsilon}
\newcommand{\ga}{\gamma}
\newcommand{\w}{\omega}
\newcommand{\bn}{\bar{n}}
\newcommand{\bq}{{\bar{q}}}
\newcommand{\hH}{\widehat{H}}
\newcommand{\hS}{\widehat{S}}
\newcommand{\hV}{\widehat{V}}
\newcommand{\cB}{{\mathcal{B}}}

\newcommand{\cP}{{\mathcal P}}
\newcommand{\bnP}{\bn\sdt{\mathcal P}}
\newcommand{\lp}{\tilde p}        
\newcommand{\vC}{\vec{C}}
\newcommand{\vO}{\vec{O}}
\newcommand{\vT}{\vec{T}}

\newcommand{\nn}{\nonumber}

\title{Combining Fixed-Order Helicity Amplitudes With Resummation Using SCET}

\ShortTitle{Combining fixed-order helicity amplitudes with resummation using SCET}

\author{Iain W.~Stewart\\
       Center for Theoretical Physics, Massachusetts Institute of Technology, Cambridge, MA 02139, USA\\
       E-mail: \email{iains@mit.edu}}

\author{\speaker{Frank J.~Tackmann}\\
        Theory Group, Deutsches Elektronen-Synchrotron (DESY), Notkestrasse 85, D-22607 Hamburg, Germany\\
        E-mail: \email{frank.tackmann@desy.de}}

\author{Wouter J.~Waalewijn\\
       Department of Physics, University of California at San Diego, La Jolla, CA 92093, USA\\
       E-mail: \email{wouterw@physics.ucsd.edu}}

\abstract{
  We discuss how to construct a simple and
  easy-to-use helicity operator basis in Soft-Collinear Effective Theory (SCET),
  for which the hard Wilson coefficients from matching QCD onto SCET are directly
  given in terms of the color-ordered QCD helicity amplitudes. This provides an
  interface to seamlessly combine fixed-order helicity amplitudes, which are the basic building
  blocks of state-of-the-art next-to-leading order calculations for multileg processes,
  with a resummation of higher-order logarithmic corrections using SCET.}

\FullConference{``Loops and Legs in Quantum Field Theory'' 11th DESY Workshop on Elementary Particle Physics \\
                 April 15-20, 2012\\
                 Wernigerode, Germany\\
                 Preprint: MIT-CTP 4409, DESY 12-180}

\begin{document}

\section{Introduction}

The production of hadronic jets in proton-(anti)proton collisions is one of the
most basic processes at hadron colliders. Key processes include those with a W, Z, or Higgs boson together
with jets. Many models for new short distance physics produce colored particles,
which would experimentally be observed as jets of energetic hadrons.  For
example, in supersymmetric models the squarks or gauginos typically decay
through a chain of (colored) particles into the lightest supersymmetric
particle. Another source of jets are the initial-state quarks and gluons taken
out of the incoming protons, which radiate before entering the hard collision.

The underlying hard process we are ultimately interested in is always accompanied by QCD corrections contributing at the various scales that are present in the process: virtual corrections to the hard process itself, real and virtual corrections from initial-state and final-state radiation, etc. The QCD radiative corrections to processes with jets are typically large. One reason is that jet selection cuts can be sensitive to additional soft and collinear emissions. In perturbation theory this sensitivity to lower scales manifests itself via large logarithms of the form $\alpha_s^n\ln^m\mu/Q$ with $m \leq 2n$. Here, $Q$ is a large scale of the order of the partonic center-of-mass energy and $\mu$ is a low scale associated with the definition of the final-state jets of the order a few tens of GeV. In the limit $\mu \ll Q$ these large logarithms degrade the fixed-order perturbation series, resulting in large perturbative uncertainties. This can be avoided (or at least alleviated) by resumming the leading (and subleading) towers of logarithms to all orders in $\alpha_s$. In practice, this results in smaller perturbative uncertainties, and hence more reliable predictions for the cross section in question.

In inclusive jet measurements, one requires a minimum number of hard jets and sums inclusively over additional emissions. In this case, one is less sensitive to lower jet scales. On the other hand, for exclusive jet cross sections, where one requires a certain fixed number of jets in the final state, one is explicitly sensitive to the jet resolution scale through the veto on additional jets. Such exclusive jet measurements play an important role in Higgs measurements and new-physics searches at the LHC.

An effective-theory framework to disentangle the relevant scales in jet production and resum the associated logarithms of ratios of these scales via renormalization group evolution (RGE) is provided by Soft-Collinear Effective Theory
(SCET)~\cite{Bauer:2000ew, Bauer:2000yr, Bauer:2001ct, Bauer:2001yt, Bauer:2002nz, Beneke:2002ph}. In SCET the soft-collinear limit of QCD is manifestly implemented at a Lagrangian and operator level using a systematic power expansion. The QCD corrections appearing at the hard-interaction scale, i.e., away from any infrared singular limits, which contain the process-specific details, are incorporated via matching from QCD onto SCET. On the other hand, the infrared-sensitive QCD dynamics below the hard-interaction scale that describes the collinear radiation within the jets and soft interactions between jets is contained in the effective theory.

\section{Resummation of Exclusive Jet Cross Sections in SCET}

Consider a process with $N$ final-state jets and $L$ leptons, photons, or other non-strongly interacting particles with underlying hard interaction
\begin{equation} \label{eq:interaction}
\kappa_a (q_a)\, \kappa_b(q_b) \to \kappa_1(q_1) \dotsb \kappa_{N+L}(q_{N+L})
\,,\end{equation}
where $\kappa_{a,b}$ denote the colliding partons, and $\kappa_i$ denote the outgoing quarks, gluons, and other particles with momenta $q_i$. The incoming partons are along the beam directions, $q_{a,b}^\mu = x_{a,b} P_{a,b}$, where $x_{a,b}$ are the momentum fractions and $P_{a,b}$ the (anti)proton momenta. We are interested in the situation where each of the partons in \eq{interaction} produces a separate identified jet and where we do not allow additional hard jets coming from a hard ISR or FSR emission. In SCET, the corresponding exclusive jet cross section can be factorized, which
leads to an expression of the form~\cite{Bauer:2002nz, Bauer:2008jx, Stewart:2009yx}
\begin{align} \label{eq:sigma}
\df\sigma &=
\int\!\df x_a\, \df x_b\, \df \Phi_{N+L}(q_a + q_b; \{q_i\})\, F(\{q_i\})
\sum_{\kappa} \tr\,\bigl[ \hH_{\kappa}(q_{a,b}, \{q_i\})\, \hS_N^\kappa \bigr] \otimes
\Bigl[ B_{\kappa_a} B_{\kappa_b} \prod_J J_{\kappa_J} \Bigr]
\,.\end{align}
Here, $\df \Phi_{N+L}(q_a+q_b, \{q_i\})$ denotes the Lorentz-invariant phase-space for the final-state particles in \eq{interaction}, where $F(\{q_i\})$ denotes the measurement made on the hard momenta of the $N$ signal jets (which in the factorization at leading order in the power expansion are approximated by the parton momenta $q_i$). Equation~\eqref{eq:sigma} does not include possible contributions from Glauber exchange.

Any dependence probing softer momenta, such as measuring jet masses or low $p_T$s, as well as the choice of jet algorithm or jet resolution variable, will affect the validity and precise form of the factorization formula \eq{sigma}. This dependence is encoded through the precise definitions of the $N$-jet soft function $\hS_N^\kappa$ (describing soft radiation), the jet functions $J_{\kappa_J}$ (describing energetic final-state radiation in a single jet) and the beam functions $B_{\kappa_{a,b}}$ (describing energetic initial-state radiation). The parton distributions of the incoming protons, $f_j$, are contained in the beam functions, which can be further factorized as $B_i = \sum_{j} {\cal I}_{i j} \otimes f_{j}$~\cite{Fleming:2006cd, Stewart:2010qs}. The beam, jet, and soft functions contain the virtual and integrated real-emission corrections in all IR-singular limits and are separately IR finite.

A jet-resolution variable with particularly simple factorization properties is $N$-jettiness~\cite{Stewart:2010tn}, which effectively provides an exclusive $N$-jet algorithm. The explicit form of \eq{sigma} for the $N$-jet cross section defined using $N$-jettiness is known~\cite{Stewart:2009yx, Stewart:2010tn, Jouttenus:2011wh}, and for $N = 0$ the resummation has been carried out to NNLL~\cite{Stewart:2010pd, Berger:2010xi}.
The resummation of exclusive jet cross sections defined using more traditional jet-clustering algorithms has been studied for example in Refs.~\cite{Ellis:2009wj, Ellis:2010rwa, Banfi:2012yh, Becher:2012qa, Tackmann:2012bt, Banfi:2012jm, Liu:2012sz}.

The remaining ingredient in \eq{sigma} is the hard function $\hH_{\kappa}(q_a, q_b, \{q_i\})$, where the sum over $\kappa \equiv \{\kappa_a, \kappa_b, \ldots \kappa_{N+L}\}$ in \eq{sigma} is over all relevant hard-interaction channels. The hard function encodes the dependence on the underlying hard interaction \eq{interaction}, including the hard virtual corrections. It is explicitly independent of the used jet definition or jet resolution variable and therefore does not depend on the precise form of the factorization.
Once the resummation for a given jet observable and a given number of jets to a certain order is known, it can in principle be applied to any desired process. In most cases the bottleneck is then to extract the required process-specific information in form of the NLO hard function from existing fixed-order calculations.

\section{Matching from QCD onto SCET With Helicity Amplitudes}

Schematically, we match onto the effective Lagrangian
\begin{equation} \label{eq:Leff}
\mathcal{L}_\mathrm{eff} = \mathcal{L}_\mathrm{SCET} \,+\, \sum_k C_k \, O_k
\,,\end{equation}
where $\mathcal{L}_\mathrm{SCET}$ is the SCET Lagrangian for soft and collinear quarks and gluons at leading order in the power expansion. The operators $O_k$ are responsible to mediate the hard interaction in \eq{interaction} by creating and destroying the appropriate number of external partons. The hard function in \eq{sigma} is given by the square of the Wilson coefficients: $H = \abs{C_k}^2$.
They are determined by requiring that the UV-renormalized amplitudes in the full and effective theories agree,
\begin{equation}
\mathcal{A}_\mathrm{QCD} \stackrel{!}{=} \mathcal{A}_\mathrm{SCET}
= \sum_k \img C_k\, \Vev{O_k}_\mathrm{SCET}
\,.\end{equation}
The Wilson coefficients explicitly depend on the UV renormalization scheme adopted for the SCET operators $O_k$, for which we use the standard dimensional regularization together with $\overline{\text{MS}}$.

As in any effective-theory setup, the IR divergences of the full theory are by construction reproduced in the effective theory and cancel in the matching. In particular, the matching coefficients $C_k$ do not dependent on the specific IR regulator used to perform the calculation. A useful choice is dimensional regularization for both UV and IR divergences, then all loop graphs in SCET are scaleless and vanish. Thus, the UV and IR divergences in SCET precisely cancel each other, and the bare matrix elements are given by their tree-level expressions. Including the counter term $\delta_O(\eps_\mathrm{UV})$ due to operator renormalization in the effective theory removes the UV divergences and leaves the IR divergences. Schematically, the SCET amplitude is thus
\begin{equation}
\mathcal{A}_\mathrm{SCET}
= \img C\cdot(\vev{O}^\mathrm{tree} + \vev{O}^\mathrm{loop})
= \img C \bigl[1 + \delta_O(\eps_\mathrm{IR}) \bigr]
\,.\end{equation}
Since the effective-theory IR divergences, $C\,\delta_O(\eps_\mathrm{IR})$, have to match those of the full theory, the matching coefficients in $\overline{\mathrm{MS}}$ are directly given by the infrared-finite part of the full-theory amplitude computed in pure dimensional regularization, $C = -\img \mathcal{A}_\mathrm{QCD}^\mathrm{fin}$.

The above simplification provided by pure dimensional regularization is well known, and was used for processes with multiple external partons for example in Refs.~\cite{Chiu:2008vv, Ahrens:2010zv, Kelley:2010fn}. Our goal is to construct a general and convenient-to-use operator basis, which lets us exploit it as much as possible. This requires to organize the possible spin and color structures, which quickly proliferate when increasing the number of external legs. It should also be easy to incorporate constraints from charge conjugation and parity invariance and to remain fully crossing symmetric. As one might expect, this can be achieved by constructing the operator basis by employing the same helicity and color decompositions as used for the QCD amplitudes. As a result, the SCET Wilson coefficients will be given directly by the IR-finite parts of the QCD color-ordered helicity amplitudes.

For each jet direction, we define two types of light-cone vectors
\begin{equation}
n_i^\mu = (1, \vec{n}_i)
\,,\qquad
\bn_i^\mu = (1, -\vec{n}_i)
\,,\end{equation}
with $n_i^2 = \bn_i^2 = 0$, $n_i\cdot\bn_i = 2$, and $\vec{n}_i = \vec{q}_i/\abs{\vec{q}_i}$ is a unit three-vector in the direction of the $i$th jet.
At leading order in the power expansion a fixed-order QCD amplitude with $N$ colored legs is matched onto operators in SCET with $N$ different collinear
fields, where the different collinear directions have to be well separated,
$n_i\cdot n_j \gg \lambda^2$ for $i\neq j$. The SCET operators are constructed out of composite fields
that are invariant under collinear gauge transformations~\cite{Bauer:2000yr,Bauer:2001ct}
\begin{equation} \label{eq:chiB}
\chi_{n,\w}(x) = \Bigl[\delta(\w - \bnP_n)\, W_n^\dagger(x)\, \xi_n(x) \Bigr]
\,,\quad
\cB_{n,\w\perp}^\mu(x)
= \frac{1}{g}\Bigl[\delta(\w + \bnP_n)\, W_n^\dagger(x)\,\img D_{n\perp}^\mu W_n(x)\Bigr]
\,.\end{equation}
Here, $\xi_n$ and $A_n$ are $n$-collinear quark and gluon fields. The $n$-collinear covariant derivative and Wilson lines are defined as
\begin{equation}
\img D_{n\perp}^\mu = \cP^\mu_{n\perp} + g A^\mu_{n\perp}
\,,\qquad
W_n(x) = \biggl[\sum_\text{perms} \exp\Bigl(\frac{-g}{\bnP_n}\,\bn\sdt A_n(x)\Bigr)\biggr]
\,.\end{equation}
The Wilson line $W_n(x)$ sums up arbitrary emissions of $n$-collinear gluons from an $n$-collinear quark
or gluon, which are $\ord{1}$ in the power counting.

Using the polarization vectors defined in the standard spinor notation (see e.g. Ref.~\cite{Dixon:1996wi}),
\begin{equation}
 \ve_+^\mu(p,k) = \frac{\mae{p+}{\ga^\mu}{k+}}{\sqrt{2} \langle kp \rangle}
\,,\quad
 \ve_-^\mu(p,k) = - \frac{\mae{p-}{\ga^\mu}{k-}}{\sqrt{2} [kp]}
\,,\end{equation}
we define a gluon field of definite helicity
\begin{equation} \label{eq:cBpm_def}
\cB^a_{i\pm} = -\ve_{\mp\mu}(n_i, \bn_i)\,\cB^{a\mu}_{n_i,\w_i\perp_i}
\,,\end{equation}
where $a$ is an adjoint color index. For example, for $n_i^\mu=(1,0,0,1)$ (and with an appropriate spinor phase convention)
we have
\begin{equation}
\ve_\pm^\mu(n_i, \bn_i) = \frac{1}{\sqrt{2}}\, (0,1,\mp\img,0)
\,,\qquad
\cB^a_{i\pm} = \frac{1}{\sqrt{2}} \bigl(\cB^{a,1}_{n_i,\w_i} \pm \img \cB^{a,2}_{n_i,\w_i} \bigr)
\,.\end{equation}
Similarly, we define a $q\bar q$-current of definite helicity
\begin{equation}
J_{ij\pm}^{\alpha\beta}
= \mp\, \varepsilon_{\mp}^\mu(\lp_i, \lp_j)\,
\frac{\mae{\bar{\chi}^{\alpha}_{n_i, -\w_i}\pm}{\gamma_\mu} {\chi^{\beta}_{n_j, \w_j}\pm}} {\sqrt{2}\braket{\lp_j\mp}{\lp_i\pm}}
\,,\qquad
\lp_i = \w_i\,\frac{n_i}{2}
\,,\end{equation}
where $\alpha$, $\beta$ are fundamental color indices, and $\lp_i$ denotes the large label momentum carried by collinear fields.
The corresponding tree-level Feynman rules are
\begin{align}
\Mae{g_{\pm}^{a}(p)}{\cB_{i\pm}^{b}}{0} &= \delta^{ab}\, \delta(\lp_i - p)
\,,\\
\Mae{q_{\pm}^{\alpha_i}(p_i)\,\bar{q}_{\mp}^{\alpha_j}(p_j)}{J_{ij\pm}^{\beta_i\beta_j}}{0}
&= \delta^{\alpha_i\beta_i}\,\delta^{\alpha_j\beta_j}\, \delta(\lp_i - p_i)\,\delta(\lp_j - p_j)
\,,\end{align}
while any other combinations of helicities vanish.

Next, we assemble the helicity fields into operators with a definite helicity structure corresponding to each external helicity configuration,
\begin{align} \label{eq:Opm_gen}
O_{\pm\pm\dotsb(\pm\dotsb;\dotsb\pm)}^{a_1 a_2 \dotsb \alpha_{i-1} \alpha_i \dotsb \alpha_{N-1} \alpha_N}
(\lp_1, \lp_2, \ldots, \lp_{i-1}, \lp_i, \ldots, \lp_{N-1}, \lp_N)
= S\, \cB_{1\pm}^{a_1}\,\cB_{2\pm}^{a_2}\dotsb J_{q\,i-1,i\pm}^{\alpha_{i-1}\alpha_i}
\dotsb J_{q'\,N-1,N\pm}^{\alpha_{N-1}\alpha_N}
\,.\end{align}
The symmetry factor $S$ takes into account the number of identical particles. For each number of positive and negative helicity gluons and quark currents, there is only one independent operator. This is because fields of the same particle type and helicity are related by interchanging their momentum labels and color indices appropriately. For example,
$O_{+-}^{a_1 a_2}(\lp_1, \lp_2) = O_{-+}^{a_2 a_1}(\lp_2, \lp_1)$ are not independent operators. To keep track of the minimal number of independent operators, we can simply order the helicity labels.

For a given $N$-parton process, the explicit form of \eq{Leff} reads
\begin{equation}
{\mathcal L}_\mathrm{eff}
= {\mathcal L}_\mathrm{SCET}
+ \sum_{\substack{\rm helicity\\\rm configurations}} \int\!\prod_{i=1}^N\df \lp_i\,
C_{+\cdot\cdot(\cdot\cdot-)}^{a_1\dotsb \alpha_N}(\lp_1, \ldots, \lp_N)\,
O_{+\cdot\cdot(\cdot\cdot-)}^{a_1 \dotsb \alpha_N}(\lp_1, \ldots, \lp_N)
,.\end{equation}
Using the above Feynman rules for the helicity fields, the resulting tree-level Feynman rule for ${\mathcal L}_\mathrm{eff}$
projects out the single Wilson coefficient that corresponds to the external helicity and color configuration,
\begin{equation}
\Mae{g_+^{a_1}(p_1) g_-^{a_2}(p_2) \dotsb q_-^{\alpha_{N-1}}(p_{N-1}) \bar q_+^{\alpha_N}(p_N)}
{\mathcal{L}_\mathrm{eff}}{0}^{(0)}_\mathrm{SCET}
 = C^{a_1 a_2\dotsb \alpha_{N-1}\alpha_N}_{+-\cdot\cdot(\cdot\cdot-)}(p_1, p_2, \ldots, p_{N-1}, p_N)
\,.\end{equation}
Using pure dimensional regularization with $\overline{\text{MS}}$ and following the same arguments as above, we thus find that to all orders in perturbation theory
\begin{equation}  \label{eq:matching_general}
C_{+-\cdot\cdot(\cdot\cdot-)}^{a_1 a_2\dotsb\alpha_{N-1}\alpha_N}(p_1, p_2, \ldots,p_{N-1}, p_N)
= -\img \mathcal{A}_\mathrm{QCD}^\mathrm{fin}\Bigl(g_+^{a_1}(p_1) g_-^{a_2}(p_2) \dotsb q_-^{\alpha_{N-1}}(p_{N-1}) \bar q_+^{\alpha_N}(p_N)\Bigr)
\,.\end{equation}

Next, to organize the color structures, we can decompose the Wilson coefficients into a complete basis of color-singlet structures,
\begin{equation} \label{eq:Cpm_color}
C_{+\cdot\cdot(\cdot\cdot-)}^{a_1\dotsb\alpha_N}
= \sum_k C_{+\cdot\cdot(\cdot\cdot-)}^k T_k^{a_1\dotsb\alpha_N}
\equiv \vT^{\dagger\, a_1\dotsb\alpha_N} \cdot \vC_{+\cdot\cdot(\cdot\cdot-)}
\,.\end{equation}
Here $\vT^{\dagger\, a_1\dotsb\alpha_N}$ is a row vector of suitable color structures that
provide a complete basis for all possible allowed color structures (but do not necessarily all have to be independent).
Up to three colored partons, there is only a single allowed color structure
\begin{equation}
\vec{T}^{\dagger\, \alpha\beta} = \bigl(\delta^{\alpha\beta}\bigr)
\,,\quad
\vec{T}^{\dagger\, ab} = \bigl(\delta^{ab}\bigr)
\,,\qquad
\vec{T}^{\dagger\, a\alpha\beta} = \bigl(T^a_{\alpha\beta} \bigr)
\,,\quad
\vec{T}^{\dagger\, abc} = \bigl(i f^{abc} \bigr)
\,,\end{equation}
while for example for $q\bq q\bq$ or $ggq\bq$ there are already two or three independent color structures,
\begin{align}
\vec{T}^{\dagger\, \alpha\beta\gamma\delta}
&= \bigl(
  \delta_{\alpha\delta}\, \delta_{\gamma\beta}\,,\, \delta_{\alpha\beta}\, \delta_{\gamma\delta}
\bigr)
\,,\qquad
\vec{T}^{\dagger\, ab \alpha\beta}
= \Bigl(
   (T^a T^b)_{\alpha\beta}\,,\, (T^b T^a)_{\alpha\beta} \,,\, \tr[T^a T^b]\, \delta_{\alpha\beta}
   \Bigr)
\,.\end{align}
Using \eq{Cpm_color}, we can rewrite \eq{Leff} in its final form
\begin{equation} \label{eq:Leff_alt}
\mathcal{L}_\mathrm{eff} = \int\!\prod_{i=1}^N\df \lp_i\,
\vO^\dagger_{+\cdot\cdot(\cdot\cdot-)}(\lp_1, \ldots, \lp_N)
\vC_{+\cdot\cdot(\cdot\cdot-)}(\lp_1, \ldots, \lp_N)
\,,\end{equation}
where the final operators are
\begin{equation} \label{eq:Opm_color}
\vO^\dagger_{+\cdot\cdot(\cdot\cdot-)}
= O_{+\cdot\cdot(\cdot\cdot-)}^{a_1\dotsb\alpha_N}\, \vT^{\dagger\, a_1\dotsb\alpha_N}
\,.\end{equation}
If we now use the same basis as in the usual color decomposition of the amplitudes
\begin{equation}
\mathcal{A}_\mathrm{QCD} \Bigl(g^{a_1}_{+}(p_1) g^{a_2}_{-}(p_2) \dotsb  \bq_{+}^{\alpha_N}(p_N) \Bigr)
= \img\sum_k \vec{T}_k^{\dagger\,a_1 a_2\dotsb\alpha_N}
A^k(1^{+}, 2^{-},\ldots,N_{\bq}^{+})
\,,\end{equation}
the $\overline{\text{MS}}$ Wilson coefficients are equal to the IR-finite parts of the color-ordered amplitudes to all orders
\begin{equation}
\vec{C}^k_{+-\cdot\cdot(\cdot\cdot-)}(p_1, p_2, \ldots, p_N)
= A^{k}_\mathrm{fin}(1^+, 2^-,\ldots,N_{\bq}^+)
\,.\end{equation}

To give a concrete and slightly nontrivial example, consider $gg q\bar q H$. There are a total of six independent helicity operators,
\begin{align} \label{eq:ggqqH_basis}
O_{++(\pm)}^{ab\, \alpha\beta}
&= \frac{1}{2}\, \cB_{1+}^a\, \cB_{2+}^b\, J_{34\pm}^{\alpha\beta}\, H_5
\,,\nn\\
O_{+-(\pm)}^{ab\, \alpha\beta}
&= \cB_{1+}^a\, \cB_{2-}^b\, J_{34\pm}^{\alpha\beta}\, H_5
\,,\nn\\
O_{--(\pm)}^{ab\, \alpha\beta}
&= \frac{1}{2} \cB_{1-}^a\, \cB_{2-}^b\, J_{34\pm}^{\alpha\beta}\, H_5
\,,\end{align}
where the factors of $1/2$ are symmetry factors. Using the above color basis for $ggq\bq$, the color decomposition of the QCD helicity amplitudes into partial amplitudes reads
\begin{align} \label{eq:ggqqH_QCD}
\mathcal{A}\bigl(g_{1\pm} g_{2\pm}\, q_{3+} \bq_{4-} H_5 \bigr)
&= \img \sum_{\sigma\in S_2} \bigl[T^{a_{\sigma(1)}} T^{a_{\sigma(2)}}\bigr]_{\alpha_3\alpha_4}
\,A(\sigma(1^\pm),\sigma(2^\pm); 3_q^+, 4_\bq^-; 5_H)
\nn\\ & \quad
+ \img\, \tr[T^{a_1} T^{a_2}]\,\delta_{\alpha_3\alpha_4}\, B(1^\pm,2^\pm; 3_q^+, 4_\bq^-; 5_H)
\,.\end{align}
The $B$ amplitudes vanish at tree-level. From \eq{ggqqH_QCD} we can read off the Wilson coefficients,
\begin{equation} \label{eq:ggqqH_coeffs}
\vC_{\pm\pm(+)}(p_1,p_2;p_3,p_4;p_5)
= \begin{pmatrix}
   A_\mathrm{fin}(1^\pm,2^\pm;3_q^+,4_\bq^-; 5_H) \\
   A_\mathrm{fin}(2^\pm,1^\pm;3_q^+,4_\bq^-; 5_H) \\
   B_\mathrm{fin}(1^\pm,2^\pm;3_q^+,4_\bq^-; 5_H) \\
\end{pmatrix}
.\end{equation}
The Wilson coefficients for the negative helicity quark current can be obtained using charge conjugation invariance,
\begin{equation}
\vC_{\lambda_1\lambda_2(-)}(p_1,p_2;p_3,p_4;p_5)
= \hV \vC_{\lambda_1\lambda_2(+)}(p_1,p_2;p_4,p_3;p_5)
\quad\text{where}\quad
\hV  =
\begin{pmatrix}
  0 & -1 & 0 \\
  -1 & 0 & 0 \\
  0 & 0 & -1
\end{pmatrix}
.\end{equation}

\section{Conclusions}

The tools to combine generic NLO calculations with a NNLL resummation are available. The required matching coefficients from matching QCD onto SCET contain the process-dependent hard virtual corrections, but do not depend on the choice of jet-resolution variable. We have shown how to explicitly construct an operator basis in SCET such that the IR-finite parts of the color-ordered partial amplitudes directly determine the hard machting coefficients. This provides a seamless interface to combine existing NLO calculations with higher-order logarithmic resummation.

It is important to note that in general the different color structures in $\vO^\dagger$ mix under renormalization. Except in the simplest cases, this already happens at NLL. Including the RGE running of the hard coefficients, the resummed cross section has the color structure
\begin{equation}
\sigma_N \sim
\vC^\dagger\cdot \widehat U_H^\dagger \,\cdot\,
\Bigl[\Bigl(B_a B_b \prod_j J_j \Bigr) \,\otimes\, \widehat S \Bigr]
\,\cdot\,\, \widehat U_H \cdot \vC
\,,\end{equation}
where the hard evolution factors $\widehat U_H$ as well as the soft function $\widehat S$ are matrices in color space, which depend on the kinematics. In practice, this means that in order to perform the resummation one requires the NLO calculation to provide access to the individual color-ordered amplitudes. For this purpose it would be important to establish a common set of conventions for the helicity and color bases.

\begin{acknowledgments}
This work was supported in part by the Office of Nuclear Physics of the U.S. Department of Energy under the Grant No. DE-FG02-94ER40818, by the Department of Energy under the Grant No. DE-FG02-90ER40546, and by the DFG Emmy-Noether Grant No. TA 867/1-1.
\end{acknowledgments}

\bibliographystyle{PoS}
\bibliography{../../pp}

\end{document}